\documentclass[twocolumn, prb, showpacs,superscriptaddress]{revtex4}
\usepackage{amsfonts}

\usepackage{amssymb}
\usepackage{graphicx}
\usepackage{dcolumn}
\usepackage{bm}
\usepackage{amsmath}

\def\OMIT#1 {{}}
\def\MEMO#1 {{}}

\begin{document}
\title{Confinement: a real-time visualization }
\author{Zi Cai}
\affiliation{Department of Physics, University of California,
San Diego, California 92093, USA}
\author{Congjun Wu}
\affiliation{Department of Physics, University of California, San
Diego, California 92093, USA}
\author{ U. Schollw\"{o}ck}
\affiliation{Department of Physics and Arnold Sommerfeld Center for
Theoretical Physics,
Ludwig-Maximilians-Universit\"{a}t M\"{u}nchen, D-80333 M\"{u}nchen,
Germany}

\begin{abstract}
Due to the mechanism of confinement, as known from quantum
chromodynamics, it is difficult to observe individual particles
carrying fractional quantum number ({\it e.g.} quark with fractional
electric charge). A condensed matter example of fractionalized
particles is spinons in quasi-one-dimensional spin systems, which
are domain walls in the background of Neel configurations carrying
spin-$\frac{1}{2}$. Using the time-evolving block decimation
algorithm, we visualize the nontrivial spinon dynamics induced by
the confine mechanism in a two-leg spin-$\frac{1}{2}$ ladder. It can
be illustrated by a simple single-particle picture of Bloch
oscillation, not only qualitatively but also quantitatively. We
propose the experimental realization and the real time detection of
the spinon dynamics in the ultra-cold boson systems of $^{87}$Rb.
\end{abstract}
\pacs{75.10 Pq, 05.70.Ln, 75.78.Fg} \maketitle

\section{Introduction}
In quantum chromodynamics, quarks are fundamental particles carrying
fractional electric charges. However, interactions between quarks
grows linearly with distance due to the mechanism of confinement
arising from the $SU(3)$ color gauge theory. Consequentially, quarks
are confined into color singlet bound states of baryons and mesons,
and thus are difficult to observe. In various condensed matter
systems, there also exist excitations carrying fractional quantum
numbers, such as the spinon and holon excitations in 1D Luttinger
liquids \cite{senechal1999}, the solitons with half-fermion charge
in one dimensional conducting polymers \cite{heeger1988}, and the
quasi-particle and hole excitations in fractional quantum Hall
systems \cite{laughlin1999}, and the monomer excitations in the
dimer model in the triangular lattice \cite{moessner2001}. In these
systems, fractional excitations are deconfined. Some of them have
been experimentally detected.

The phenomenon of confinement  emerges in quasi-1D strongly
correlated systems such as spin ladders.
The ``quark'' in this case is
known as spinon, which is the domain wall interpolating between
different ordered regions and usually carries fractionalized spin
(spin-$1/2$). Take an antiferromagnetic spin chain with Ising-like
exchange anisotropy for an example, the ground state in this case
would be a Neel state with two-fold degeneracy, while the spinon can
be considered as the elementary excitation separating these two
degenerate ground states with opposite staggered magnetization. In a
single spin chain, no confinement occurs because a pair of spinons
could be separated as far from each other as possible without
costing energy. For a two-leg ladder system, even an infinitesimal
interchain coupling would induce a potential which increases
linearly with distance between the
spinons \cite{Dagotto,Shelton,Greiter,Bhaseen}. Therefore a pair of
fractionalized spinon excitations would be confined into an integral
spin-1 excitation: a magnon.  Recently, a finite temperature
confined-deconfined crossover has been observed in neutron
scattering experiments for a weakly coupled ladder material
CaCu$_2$O$_3$\cite{Lake}. The signature of the confinement of spinons
can be observed from the energy absorption spectrum for spin-flips
at various wavevectors. In this paper, we reexamine this old concept
from a novel perspective, which enables us to visualize the
confinement directly through the non-equilibrium dynamics in a cold
atom system.

Due to the low dissipation rate and the long coherence times,
ultracold atoms in optical lattices have opened exciting
possibilities for studying non-equilibrium quantum dynamics of
many-body systems \cite{Kinoshita,Winkler,Trotzky1,Hassanieh,Hung,
Chen2011,Kasztelan}. On the other hand, it also provides a perfect
platform to reexamine classic concepts in condensed matter or
particle physics from a different perspective. One example is the
spin-charge separation, which plays a central role in strongly
correlated systems\cite{Lieb,Baskaram}. Recently, Kollath $et$
$al$\cite{Kollath} have used the adaptive t-DMRG\cite{Daley} to
study the time evolution of a 1D fermionic Hubbard model in real
time, and observed the splitting of local perturbation into separate
wave packets carrying charge and spin. In this paper, we study the
time evolution of a pair of spinons in a two-leg spin ladder model,
and find that the confinement mechanism would lead to nontrivial
dynamics of the spinons, which enables us to visualize confinement
in real time.

\begin{figure}[htb]
\includegraphics[width=0.85\linewidth]{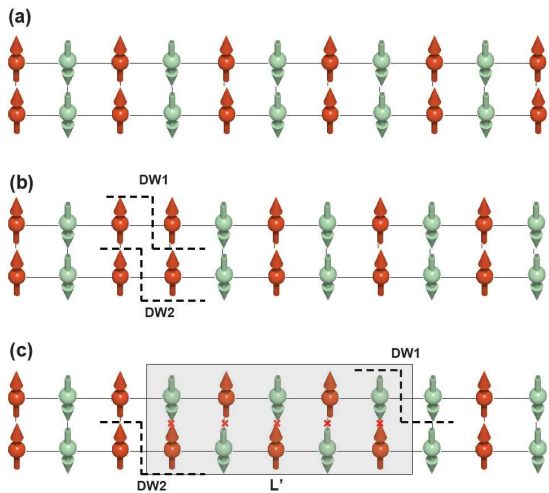}
\caption{Confinement on the two-leg spin-$\frac{1}{2}$ ladder with
Ising anisotropy described by Eq.(\ref{eq:ham}). (a) The classic
ground state with Neel ordering along the leg and ferromagnetic
ordering along the rung. (b) The classic configuration with
$spinon_1$ and $spinon_2$ located at the same position. (c) The
mismatch of the locations of $spinon_{1,2}$ gives rise to the energy
cost linearly increasing with the mismatch length. } \label{fig1}
\end{figure}

\section{Confinement in a two-leg spin ladder}
Our departure point is a two-leg spin-$\frac{1}{2}$ ladder with easy
axis anisotropy along the $z$-axis defined as
\begin{eqnarray}
H&=&\sum_{i,a} J_{a, \parallel} (S_{i,a}^+
S^-_{i+1,a}+h.c)+J^{z}_{a, \parallel} S_{i,a}^z
S_{i+1,a}^z \nonumber \\
&-& J^{z}_{\perp} S_{i,1}^z S_{i,2}^z,
\label{eq:ham}
\end{eqnarray}
where $a=1,2$ is the leg index; $\parallel$ and $\perp$ denote the
couplings along the leg and across the rung, respectively. The
couplings within legs are antiferromagnetic with the Ising
anisotropy ($J^{z}_{a,\parallel}>2 J_{a,\parallel}>0$). The rung
coupling is chosen as Ising-like and ferromagnetic. The experimental
realization of Eq.(\ref{eq:ham}) in ultra-cold atom systems will be
discussed later. Below we will focus on the case of
$J_{1,\parallel}\neq J_{2,\parallel}$ which gives rise to the
nontrivial dynamics induced by spinon confinement.

Because of the Ising anisotropy and the ferromagnetic rung coupling,
the ground state is a Neel state within each leg and with the spins
of each rung aligned as shown in Fig. \ref{fig1}
(a)\cite{Giamarchi}. Introducing two spinons in each leg
respectively located at the same position cost the domain wall
energy $E_b=\frac{1}{2} (J^z_{1,\parallel}+J^z_{2,\parallel})$ as
shown in Fig. \ref{fig1} (b). The separation of $spinon_1$ and
$spinon_2$ will lead to a mismatch within the rungs between
$spinon_1$ and $spinon_2$. This induces a finite string tension
between spinons that grows with distance, leading to the
confinement\cite{Bhaseen}, as shown in Fig. \ref{fig1} (c). The
transverse spin couplings $J_{a,\parallel}$ generate spinon
dynamics. In the absence of the rung coupling, two spinons move
independently with velocities $v_a \propto J_{a,\parallel}$
\cite{Gobert}. In the case of $J_{1,\parallel} \neq J_{2,\parallel}
$, $spinon_1$ and $spinon_2$ propagate with different velocities,
thus becomes apart eventually. The effective potential between
spinons is known as \cite{Greiter}:
\begin{eqnarray}
V(r)=k |\frac{r}{a}|, \quad k\propto J^{z}_{\perp} \langle G|
S_{i,1}^zS_{i,2}^z| G\rangle,
\end{eqnarray}
where $r$ is the average distance between these two spinon; $a$ is
the lattice constant;  $\langle G| S_{i,1}^zS_{i,2}^z |G\rangle$ is
taken over the ground state.

\begin{figure}[htb]
\includegraphics[width=0.9\linewidth]{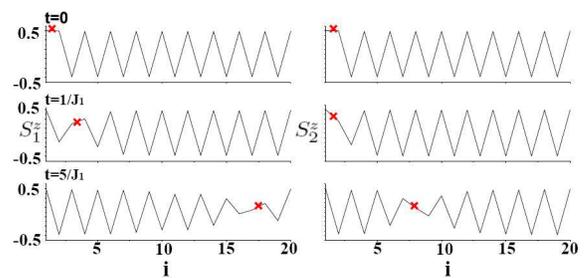}
\caption{Time evolution of the spinon dynamics with $J^z_\perp=0$
with the spinon positions marked by the red cross. The spinons in
the 1st (left) and 2nd (right) leg move independently.} \label{fig2}
\end{figure}

\section{Confinement-induced dynamics of spinons}

In our numerical simulation, we use the open boundary condition
(OBC) and prepare the initial state as both spinons located at the
left end of the ladder, i.e., between the first and second sites.
Right after the beginning of the evolution, the spinons will be
rebounded by the left boundary and propagate towards the left end
with a velocity proportional to $J_{a,\parallel}$. The length of the
ladder is set to $L=20$, long enough for the time scales simulated to exclude
finite-size effects. Utilizing the time-evolving block
decimation (TEBD) algorithm \cite{Vidal}, we study the time
evolution of the many-body wavefunction from the initial state.
Total $S^z$ conservation is used to reduce the computational effort.
In the course of real time evolution we take the truncation
dimension $\chi=80$ and time step $\Delta\tau=0.05$. The convergence
is checked by taking larger $\chi$.

\begin{figure}[htb]
\includegraphics[width=8.5cm]{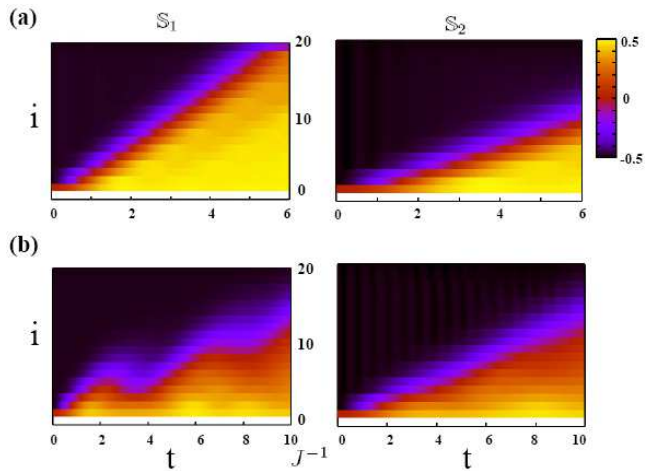}
\caption{The time evolution of the spatial distribution of
$\mathbb{S}_{i,a=1,2}$ for (a) $J^{z}_{\perp}=0$ and (b)
$J^{z}_{\perp}=J$. They exhibit the propagations and oscillations of
the spinons in the first (left) and second (right) legs. }
\label{fig3}
\end{figure}

The calculated time evolution for the case of $J^{z}_{\perp}=0$ is
shown in Fig. 2 in which two spinons are decoupled. The parameter
values are taken as $J_{1,\parallel}=J, J_{2,\parallel}=0.5J$, and
$J^{z}_{1,\parallel}= J^z_{2,\parallel}=5J$. To visualize the spinon
dynamics, we present the time-evolution of the configuration of the
expectation value of $S_z$. The spinons are located at the bonds
connecting two sites with the same sign of $S_z$. During the
evolution, two spinons become separated due to their different
velocities determined by $J_{a,\parallel}$. For a clear presentation
of the spinon, we define the rectified  magnetization
$\mathbb{S}_{i,a}$ for each leg as
\begin{equation}
\mathbb{S}_{i,a}=-S^z_{i,a}\times (-1)^i.
\label{rectify}
\end{equation}
The location of $spinon_{1}$ or $spinon_2$ is determined at the bond
across which $\mathbb{S}_{i,a}$ changes the sign. The time evolution
of the spatial distributions of $\mathbb{S}_{i,a=1,2}$ at
$J^z_\perp=0$ is depicted in Fig. \ref{fig3} (a). Both legs exhibit
the propagation of spinons at uniform speeds. The speed of the first
leg is larger than the that of the second one because
$J_{1,\parallel}>J_{2,\parallel}$.

Now we turn on the coupling along the rung, a confinement potential
emerges between two spinons. It acts like a ``constant'' force
connecting these two spinons. If we view the spinons as particles,
the dynamics of the spinons can be qualitatively understood from a
simple two-body picture as follows. Two particles with different
initial velocities and masses move in a 1D lattice system and
interact with each other via a linear potential. Each particle
oscillates around the center of mass (COM) of the system, which
continues to do uniform linear motion until it reaches the boundary.
This simple picture could be verified qualitatively by our numerical
simulation of the spinon dynamics. The time evolution of spatial
distribution of $\mathbb{S}_{i,a=1,2}$ at $J^{z}_{\perp}=J$ is
presented at Fig. \ref{fig3} (b). An oscillation of the $spinon_1$
is observed, while that of $spinon_2$ is not as clear as $spinon_1$.
The effective mass of $spinon_2$ is larger than that of $spinon_1$,
thus its oscillation amplitude is too small to be visible.

\begin{figure}[htb]
\includegraphics[width=0.8\linewidth]{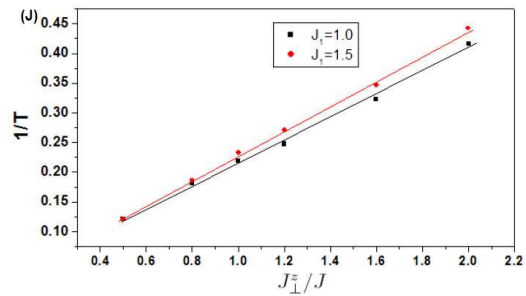}
\caption{The relation of $1/T$ v.s. $J^{z}_{\perp}$ with parameter
values of $J_{2,\parallel}=0$, and $J_{1,\parallel}=J$ (squares) or
$1.5J$ (circles).} \label{fig4}
\end{figure}

To further verify the two-particle picture quantitatively, we choose
$J_{2,\parallel}=0$, which means $spinon_2$ is localized in its
initial position at the boundary. In this case, the two-body problem
reduces to a one-body problem of $spinon_1$ in a lattice system
under a static magnetic field, which is provided by the rectified
spins in the second leg.
The time evolution of a spinon within a ferromagnetic spin chain
under a constant magnetic field has been studied previously, the
dynamics turns out to be a perfect Bloch oscillation \cite{Cai1}. In
our case, the confinement 'constant' force provided by the static
$spinon_2$ plays a similar role of the constant electric field
($E\propto J^{z}_{\perp}$), and we expect a similar Bloch
oscillation dynamics. It implies that the frequency of the
oscillation of $spinon_1$ should be proportional to the interaction
strength: $1/T \propto J^{z}_{\perp}$, where $T$ is the oscillation
period. This relation can be verify numerically, as shown in Fig.
\ref{fig4}, where we can find a linear relation between $1/T$ and
$J^{z}_{\perp}$ for both $J_{1,\parallel}=J$ and $1.5J$. Simulation times
were long enough to observe the stability of the oscillation phenomenon.

In a perfect Bloch oscillation, the frequency is determined by the
strength of the external field ($J^{z}_{\perp}$ in our case), and
should be independent of the band width ($J_{1,\parallel}$ in our
case). However, we find that the oscillation frequency not only
depends on $J^{z}_{\perp}$, but is also slightly dependent on
$J_{1,\parallel}$, especially for large values of $J^{z}_{\perp}$.
There are two possible factors contributing to the imperfection of
Bloch oscillation. One is the boundary effect. Considering the
quantum nature of spinons (wave packets), the boundary begin to
influence the dynamics even before the centers of spinons reach the
boundary. For larger $J^{z}_{\perp}$, the amplitude of the Bloch
oscillation $A$ is smaller ($A\propto
J_{1,\parallel}/J^{z}_{\perp}$), which indicates the motion of the
$spinon_1$ is confined to the boundary, therefore the boundary
effect becomes more obvious. This can explain why the deviation
becomes larger for bigger $J_{1,\parallel}/J^{z}_{\perp}$. The other
reason comes from deviations from the linear relations of Eq. (2),
which can be considered as a zero-order approximation of the exact
interaction between spinons. Actually, on a long length scale, it is
possible that higher order nonlinear modifications such as $1/|r|$
emerge due to the quantum nature of spinons \cite{Bhaseen}.

\section{Experimental realization}
\begin{figure}[htb]
\includegraphics[width=0.8\linewidth]{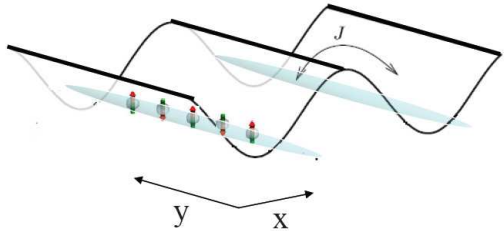}
\caption{Sketch picture for the experimental setup of
our two-leg spin ladder system.}
\label{fig5}
\end{figure}

For an experimental realization of our two-leg spin ladder system in
an ultracold atomic superlattice system, one first has to create
Neel-ordered chains along the $x$-direction \cite{Trotzky}, for
example by using $^{87}$Rb atoms with two internal spin states
$|F=1,m_F=\pm 1\rangle$. The necessary superexchange interactions of
the ultra cold atoms can be achieved by the use of a superlattice
system \cite{Trotzky,Chen,Cirac,Duan,Kuklov}. In the optical
lattice, a typical energy scale for the superexchange coupling $J$
is in the order of kHz, which implies a $ms$ time-scale  real-time
dynamics in our case. By modifying the bias between neighboring
lattice sites, both ferromagnetic and antiferromagnetic
superexchange interaction can be realized \cite{Trotzky}; the
orientation of the setup would be as shown in Fig. \ref{fig5}. The
two-leg ladder structure is achievable by applying another laser
along the $y$-direction to construct a second superlattice
structure. In addition, the easy-axis anisotropy is realizable by
tuning the interaction
$U_{\uparrow\uparrow}=U_{\downarrow\downarrow}\neq
U_{\uparrow\downarrow}$ \cite{Duan,Kuklov} or by a periodically
modulated lattice \cite{Chen}. Local addressing techniques as
developed recently \cite{Weitenberg} would be used to create domain
walls, e.g. by a global spin-flip on one half of the system, as well
as for the read-out procedure. These techniques, while advanced, are
now experimentally established. The arguably most difficult part of
the experiment would be to adjust the exchange interaction
separately along two different legs. As mentioned above, the
exchange interaction can be adjusted by superlattices
\cite{Trotzky}, but to do this separately on legs 1 and 2, an
additional superlattice structure is needed. In fact,
antiferromagnetic spin chains have already successfully been
simulated, albeit in a different context, namely the Bose-Hubbard
model in a tilted optical lattices, where a magnetic domain wall has
been observed \cite{Simon}.

\section{Discussion and conclusion}

In the context of condensed matter physics, the study of domain wall
motion in magnetic nanowires has attracted considerable theoretical
and experimental attentions recently due to its potential industrial
applications. Manipulation and control the domain wall dynamics in
magnetic nanowires is known to play an important role in
nanomagnetism \cite{Parkin,Beach,Klaui}. Most of the theoretical
studies in this field are based on the Landau-Lifshitz-Gilbert (LLG)
equation\cite{Schryer,Wang}. Our work is different from previous
work in three aspects: (i) we provide a quasi-exact calculation of the
dynamics of the strongly correlated system by TEBD instead of a
semi-classic mean-field approximation; (ii) different from condensed
matter system where the domain wall dynamics is closely related with
dissipation rate, there is little energy dissipation in our system;
(iii) we account for the strong confined interaction between  domain
walls, which leads to nontrivial dynamics.

Whether in condensed matter or ultracold atom physics, the control
and manipulation of the particles or quasiparticles in the quantum
system are of central interest. In this paper, confinement is
visualized in real space and its properties are studied via a
non-equilibrium process, showing a possible way to investigate and
manipulate the motion and interaction of quasiparticles in a
dissipationless strongly correlated system. Our work could be
helpful for the study of phenomena involving the domain wall
dynamics, such as Walker breakdown \cite{Schryer} and current-driven
domain wall motion \cite{Klaui}.

We are grateful to S. Trotzky for intensive discussions about
the experimental realization. Z.C. thanks Lei Wang for numerical
help.
This work was supported in part by the NBRPC (973 program) 2011CBA00300
(2011CBA00302).
Z.C. and C.W. are supported by NSF DMR-1105945, and the AFOSR-YIP
program. U.S. is supported by DFG through FOR801.

\end{document}